\documentstyle[12pt]{article}

\begin{document}
\begin{center}
{\Large {\bf Symplectic Quantization of Open Strings and 
Noncommutativity in Branes}}\\
\vspace{1cm} 

{\large Nelson R. F. Braga$^{a}$  and Cresus F. L. Godinho$^{b}$}  \\
 
\vspace{0.5cm}

{\sl
$^a$ Instituto de F\'{\i}sica, Universidade Federal
do Rio de Janeiro,\\
Caixa Postal 68528, 21945-970  Rio de Janeiro, Brazil\\[1.5ex]

$^b$ Centro Brasileiro de Pesquisas F\'{\i}sicas, Rua Dr Xavier Sigaud 150,\\
 22290-180,
Rio de Janeiro RJ Brazil }
\end{center}
\vspace{1cm}

\abstract  

We show how to translate boundary conditions into constraints in the symplectic
quantization method by an appropriate choice of generalized variables.
This way the symplectic quantization of an open string attached to a brane
in the presence of an antisymmetric background field reproduces the  
non commutativity of the brane coordinates.

\vskip3cm
\noindent PACS: 11.15 , 03.70

\vspace{1cm}

\vspace{1cm}

\noindent braga@if.ufrj.br; godinho@cbpf.br

\vfill\eject

The interest in studying space time with non commutative structure is not new
but has widely increased after important results from 
string theory. A general discussion and an important list of references can be 
found in\cite{SW}. 
Among these results is the remarkable appearance of non commutative coordinates
orthogonal to merging Branes\cite{Wi}. Also the coordinates of
string endpoints, that means the D-brane\cite{Po} world volumes, are noncommutative
when an antisymmetric constant field is present\cite{HV,CH1}.

Strings  attached to branes involve mixed (combination of Dirichlet
and Neumann) boundary conditions. This  makes the quantization procedure more subtle. 
The standard canonical commutation relations can not be imposed as quantum commutators
as they are not consistent with the boundary conditions.
This situation is analogous to that of systems with constraints, where one must build
up commutators that are consistent with them. The difference is that
the boundary conditions, in the form that comes from the functional variation of 
the string action, involve velocities. So, they do not correspond to standard Dirac 
constraints.  If we apply directly standard Dirac procedure to the string 
action we would not find the boundary conditions as constraints.
Nevertheless, it has been recently shown in refs. \cite{CH2,AAJ,JS} that 
it is possible to use the Dirac procedure as long as one rewrites
the boundary conditions in terms of phase space variables and introduces them
as constraints. Then the interesting result of the non commutativity of the string
end points emerges.

We will show here that if instead of Dirac procedure, symplectic quantization scheme 
is considered, the boundary conditions arise directly as constraints  by means of a
discretization  of the string worldsheet spatial coordinate. 
We mean: starting with a discretized version of the string action and choosing 
appropriate phase space coordinates the boundary conditions will emerge directly as 
constraints (zero modes of the symplectic matrix) in the quantization procedure. 
They do not have to be introduced by any additional mean but rather they show up by just 
following the standard symplectic algorithm.
We will see that the discretization, also used in ref. \cite{AAJ} but as a tool 
for avoiding singularities in the inversion of the Dirac matrix,  will play here 
an additional  role. It will make it possible to define the phase space variables for
the string endpoints in such a way to capture the boundary conditions in the 
symplectic matrix and find the coordinate commutators.

The symplectic quantization was proposed in \cite{FJ} as an alternative to
Dirac procedure for constrained systems.  In this approach the calculation of  
commutators for first order Lagrangians is much more straightforward. 
The role of constraints and gauge symmetries
in this  quantization framework has bee discussed in \cite{BW,Wo,Mo}. 
In the standard form, symplectic quantization is not sensible to 
boundary conditions and would not reproduce the peculiar non 
commutativity of the string end points in the presence of an antisymmetric 
constant field. We will show here how to find a particular choice of 
symplectic variables such that the boundary conditions show up naturally as constraints. 
We do not have to impose them 

This way we find a straightforward way of building  up the coordinate commutators 
consistent with the mixed boundary conditions and thus  reproduce the non 
commutativity at the string end points.

In the symplectic quantization one takes as the starting point a Lagrangian 
that is first order in time derivatives. 

\begin{equation}
\label{FO}
L^0 = a_k^0 ( q ) \dot q_k - V( q ) 
\end{equation}

\noindent where $q_k$ are the generalized coordinates of the system. 
(If the original Lagrangian is of higher order we can rewrite it in the 
form of eq. (\ref{FO})  by simply  introducing new symplectic variables.)

Then we build up the symplectic matrix

\begin{equation}
\label{f0}
f^0_{kl} \, = \, { \partial a^0_l \over \partial q_k} -
{ \partial a^0_k \over \partial q_l}\,\,.
\end{equation}

\noindent If this matrix is non singular we can define the commutators of 
the quantum theory (if there is no ordering problem for the corresponding 
quantum operators) as

\begin{equation}
\lbrack A (q) , B(q) \rbrack \,=\, { \partial A \over \partial q_k} (f^0)^{-1}_{kl} 
{ \partial B \over \partial q_l}
\end{equation}

\noindent If the matrix is singular we find the zero modes $v^{\alpha}_l (q)\,$ 
 that satisfy: $ f^0_{kl} v^{\alpha}_l \,=\,0\,$ and the corresponding constraints:

\begin{equation}
\Omega^\alpha \,=\,v^{\alpha}_l {\partial V \over \partial q_l } \approx 0
\end{equation}

\noindent Then we introduce new variables $\lambda^{\alpha}$ and take as the new 
Lagrangian

\begin{equation}
L^1 = a_k^0 ( q ) \dot q_k + {\dot\lambda}^{\alpha} \Omega^{\alpha} - V( q ) \,\equiv \,
a_r^1 ( {\tilde q} ) {\dot {\tilde q}}_r - V( q )
\end{equation}

\noindent where we introduced the new notation for the extended variables:
 $ {\tilde q}^r \,=\,  ( q^k , \lambda^\alpha ) $. We find now the new matrix 
$ f^1_{rs}$   

\begin{equation}
f^1_{rs} \, = \, { \partial a^1_s \over \partial {\tilde q}_r} -
{ \partial a^1_r \over \partial {\tilde q}_s}\,\,.
\end{equation}

If the matrix $f^1$ is not singular as will be the case in this article, we define 
the quantum commutators as 

\begin{equation}
\lbrack A (\tilde q ) , B( \tilde q ) \rbrack \,=\, 
{ \partial A \over \partial {\tilde q}_r} 
(f^1)^{-1}_{rs} 
{ \partial B \over \partial {\tilde q}_s}\,\,.
\end{equation}

\noindent If the matrix is still singular the process is repeated, building up Lagrangians 
$ L^2 , L^3  $, ... until no more zero modes are found.
 
\bigskip

The action for an open string coupled to an antisymmetric field on the brane
can be taken as \cite{SW}

\begin{equation}
\label{1}
S \,=\,
{ 1 \over 4\pi \alpha^\prime }
\int_{\Sigma} d^2\sigma \Big(
\eta_{i j } \partial_a X^i \partial_b X^j g^{ab}
\,+ \, \epsilon_{ab}{\cal B}_{ij} \partial_a X^i \partial_b X^j 
 \Big)
\end{equation}

\noindent where $X^i $ are the spacetime string coordinates, ${\cal B}_{ij}$
is the antisymmetric field, and $\Sigma$ is the string world sheet where 
 $g^{ab} = (1,-1) $,  $\epsilon_{01} = -1$. We can take  
$\sigma_1 \equiv \tau \,,\,\sigma_2\equiv \sigma\,$
and represent, as usual, the boundary (string endpoints) as
$\sigma = 0\,,\,\pi\,$. The boundary conditions satisfied by the string coordinates are
then

\begin{eqnarray}
\label{e1}
\Big( \partial_\sigma X^i &-&
 {\cal B}_{ij} \partial_\tau X^j \Big)\vert_{\sigma = 0}  \,=\,0\nonumber\\
\Big( \partial_\sigma X^i &-& {\cal B}_{ij} \partial_\tau X^j \Big)\vert_{\sigma = \pi}
\,=\,0
\end{eqnarray}

Let us see how these conditions show up as equations of motion  
by writing down a discretized version of the  Lagrangian associated with (\ref{1}).
This analysis will show us how to implement boundary conditions in the symplectic
quantization. 
Dividing the interval $0 < \sigma < \pi $ in $N$ intervals of length $\epsilon$ 
and introducing the coordinates of the endpoints of the intervals as 
$X^i_0\,,\,X^i_1\,,...,\,X^i_N$ we find

\begin{eqnarray}
\label{L}
L (\epsilon ) &=& {\epsilon \over 4\pi \alpha^\prime } \,\lbrace \,
( \partial_\tau X^i_0 )^2  
\,+ \, ( \partial_\tau X^i_1)^2  \,+\,...\,+
(\partial_\tau X^i_N )^2 \, -\, { ( X^i_1 - X^i_0 )^2 \over \epsilon^2}\nonumber\\
& & \nonumber\\
&-&
{ ( X^i_2 - X^i_1 )^2 \over \epsilon^2}\,-...\,-\, 
{ ( X^i_N - X^i_{N-1} )^2 \over \epsilon^2}
\Big)\,-\, 2 {\cal B}_{ij} \partial_\tau X^i_0 {\Big( X^j_1\,-\, X^j_0 
\Big)\over \epsilon}  
\nonumber\\ 
& &\nonumber\\
&-& 2 {\cal B}_{ij}  \partial_\tau X^i_1 {\Big( X^j_2\,-\, X^j_1 \Big) \over \epsilon}  
\,-\,... - 2 {\cal B}_{ij}  \partial_\tau X^i_{N - 1} {\Big( X^j_N \,-\, X^j_{N-1} \Big)
 \over \epsilon}  \,\,
 \rbrace\,.\nonumber\\
\end{eqnarray}

\noindent using the notation: $(u^i)^2 \equiv \eta_{ij} u^i u^j \,$.

The equations of motion for respectively  $X^i_0\,,X^i_n\,,\,X^i_N $ 
(with $ 1 \le n \le N-1$) are 

\begin{eqnarray}
\label{EM}
\epsilon \partial^2_\tau X^i_0 &-& {X^i_1 - X^i_0 \over \epsilon}\,-\,
 {\cal B}_{ij} \partial_\tau X^j_1\,+\,
2 {\cal B}_{ij} \partial_\tau X^j_0\,=\,0\,\,,
\nonumber\\
\epsilon \partial^2_\tau X^i_n &-& {X^i_{n+1} - X^i_n \over \epsilon}
\,+\, {X^i_n - X^i_{n-1} \over \epsilon}\,-\,
{\cal B}_{ij} \partial_\tau X^j_{n-1}\,-\,
{\cal B}_{ij} \partial_\tau X^j_{n+1}\,+\,
2 {\cal B}_{ij} \partial_\tau X^j_n\,=\,0\,,\nonumber
\\
\epsilon \partial^2_\tau X^i_N &+& {X^i_N - X^i_{N-1} \over \epsilon}\,-\,
 {\cal B}_{ij} \partial_\tau X^j_{N-1}\,\,\,.
\end{eqnarray}

When we take the limit $\,\epsilon \rightarrow 0 \,$ and consider that 
$X^i_1 \rightarrow X^i_0\,$ and $X^i_{N-1} \rightarrow X^i_N\,$ the equations for 
$\,X^i_0 \,$ and $\, X^i_N\,$ give the  open string boundary conditions of eq. (\ref{e1}).

The equations for points $X^i_n \,$  give no 
contribution at order zero in
 $\epsilon$ but to order one  in $\epsilon$ (dividing by $\epsilon$ and then  taking 
the  limit $\epsilon \rightarrow 0 $)  they take the form  of the standard equation 
of motion for the string coordinates

\begin{equation}
\label{e2}
\partial^2_\tau X^i_n \,- \, \partial^2_\sigma X^i_n \,=\,0\,.
\end{equation}

It is important to remark that if we consider some coordinate $X^i_n$ with fixed 
$n$ and take the limit $\epsilon \rightarrow 0 $ this coordinate would tend to the 
end point $\sigma = 0$. Thus when we want to look at points inside the string 
(with a finite distance to the boundary) we must look at some $X^i_n$ but increase $n$ 
when we take the limit $\epsilon \rightarrow 0$ in order to look at a fixed point.

This analysis of the equations of motion shows us that  in the discretised version
both boundary conditions and string  equations of motion  show up in
the set of generalized equations of motion but {\bf at different orders in the 
discretization parameter $\epsilon$}. 
This will help us to find an appropriate definition of the symplectic variables
such that the boundary conditions lead to zero modes in the symplectic matrix.

The standard way of writing the Lagrangian $L$ of eq. (\ref{L}) in a first order form 
would be to introduce  the conjugate momenta $\Pi$  as symplectic variables and
eliminate the second order time derivatives.
The conjugate momenta associated with the string coordinates are

\begin{eqnarray}
\Pi_0^i &=& { 1 \over 2\pi \alpha^\prime} \lbrace  \epsilon \partial_\tau X^i_0 
\,-\, {\cal B}_{ij} \Big( X^j_1 - X^j_0 \Big) \,\rbrace \,,\nonumber\\
\Pi_n^i &=& { 1 \over 2\pi \alpha^\prime} \lbrace \epsilon \partial_\tau X^i_n
\,-\, {\cal B}_{ij} \Big( X^j_{n+1} - X^j_n \Big) \,
\rbrace\,,\nonumber\\
\Pi_N^i &=& { 1 \over 2\pi \alpha^\prime} \epsilon \partial_\tau X^i_N 
\,,
\end{eqnarray}

\noindent where again $1 < n < N$. The asymmetry in the equations for 
the $0$ and $N$ indices is just a consequence of the choice of right derivatives
but it becomes irrelevant in the $\,\epsilon \rightarrow 0 \,$ limit, as we saw
in case of the equations of motion.

We have just learned that the Euler Lagrange equations lead both to
the boundary conditions and the usual string equations of motion but at different 
orders in the parameter  $\epsilon$. Inspired by this fact we will
incorporate  in the symplectic quantization procedure the 
string boundary conditions of eq. (\ref{e1}) by introducing as new symplectic
variables not the canonical momenta but rather variables $P$ 
that also involve different orders of $\epsilon$:

\begin{equation}
\label{SV}
P_0^i \,=\,{\Pi_0^i \over \epsilon}\,\,\,,\,\,\,
P_1^i \,=\, \Pi_1^i \,\,\,\,,...\,\,\,,
P_n^i \,=\, \Pi_n^i \,\,\,\,,\,...\,\,,
P_{N-1}^i \,=\,\Pi_{N-1}^i \,\,,\,\,
P_N^i \,=\,{\Pi_N^i \over \epsilon}\,\,.
\end{equation}

\noindent The symplectic first order Lagrangian is then

\begin{equation}
L^0 \,=\,\eta_{ij} \Big( \epsilon P_0^i  \partial _\tau X^j_0,+
 \,P_1^i  \partial _\tau X^j_1\, + ...+ P_n^i  \partial _\tau X^j_n \,+\,...+
P_{N-1}^i  \partial _\tau X^j_{N-1} \,+\,
\epsilon P_N^i  \partial _\tau X^j_N \,\Big)\,+ V
\end{equation}

\noindent where

\begin{eqnarray}
V &=& - {1\over 4\pi\alpha^\prime} 
\Big( {(X^i_1 - X^i_0 )^2 \over \epsilon }\,+\, 
{ (X^i_2 - X^i_1 )^2 \over \epsilon }\,+\,...\,+\,
{ (X^i_N - X^i_{N-1})^2 \over \epsilon }\,\Big)\nonumber\\
&-& {\epsilon \over 4\pi\alpha^\prime}\Big[\,\, \Big( 2\pi \alpha^\prime P_{0i} \,+\,
{\cal B}_{ij} {\big( X_1^j - X_0^j\Big) \over \epsilon} \,\Big)^2 
\,+\, \Big(
 {2\pi \alpha^\prime \over \epsilon }  P_{1i} 
\,+\,{\cal B}_{ij} {\big( X_2^j - X_1^j\Big) \over \epsilon} \,
\Big)^2 \,
 \,...\nonumber\\
&+& \Big(  {2\pi \alpha^\prime \over \epsilon }  P_{N-1}^i 
+\,{\cal B}_{ij} {\big( X_N^j - X_{N-1}^j \big) \over \epsilon}\Big)^2 \,
+\, \Big( 2\pi \alpha^\prime P_{Ni} \Big)^2 \,\, \Big]
\end{eqnarray}

We build up the symplectic matrix $f_0$ of eq. (\ref{f0}) 
with our coordinates $X\,,\,P\,$

\begin{equation}
\label{matriz}
\bordermatrix{&X_0^j&P_0^j&X_1^j&P_1^j&...&X_n^j&P_n^j&...&X_N^j&P_N^j\cr
X_0^i&0&-\epsilon\delta^{ij}&0&0&...&0&0&...&0&0\cr
P_0^i&\epsilon\delta^{ij}&0&0&0&...&0&0&...&0&0\cr 
X_1^i&0&0&0&-\delta^{ij}&...&0&0&...&0&0\cr
P_1^i&0&0&\delta^{ij}&0&...&0&0&...&0&0\cr
...&...&...&...&...&...&...&...&...&...&...\cr
X_n^i&0&0&0&0&...&0&-\delta^{ij}&...&0&0\cr
P_n^i&0&0&0&0&...&\delta^{ij}&0&...&0&0\cr
...& ...&...&...&...&...&...&...&...&...&...\cr
X_N^i&0&0&0&0&...&0&0&...&0&-\epsilon\delta^{ij}\cr
P_N^i&0&0&0&0&...&0&0&...&\epsilon\delta^{ij}&0\cr}
\end{equation}

Now we see the reason for  introducing the symplectic variables (\ref{SV}):
in the limit $\epsilon \rightarrow 0$, when the string theory is recovered,
the  matrix (\ref{matriz}) becomes singular with the zero modes 

\begin{equation}
v_1^i \,=\,\pmatrix{u^i\cr0\cr0\cr0\cr...\cr0\cr0\cr}
\hskip 2cm
v_2^i\,=\,\pmatrix{0\cr u^i\cr0\cr0\cr...\cr0\cr0\cr}
\hskip 2cm
v_3^i\,=\,\pmatrix{0\cr0\cr0\cr0\cr...\cr u^i\cr0\cr}
\hskip 2cm
v_4^i\,=\,\pmatrix{0\cr0\cr0\cr0\cr...\cr0\cr u^i\cr}\,\,,
\end{equation}

\noindent where $u^i$ is a column vector with value one in the line $i$ and zero elsewhere
and all the zeros in the above matrices are actually column vectors in the space 
corresponding to the index $i\,$ with all elements vanishing.
These zero modes are in principle associated with constraints 
$\Omega_p^i \,=\, {\partial V\over \partial v_p^i} $.

We find for the first one:

\begin{equation}
\Omega_1^i \,=\,
{( X^i_1 - X^i_0 ) \over \epsilon } \, - \, 2 \pi \alpha^\prime\,
 M^{-1}_{ij}{\cal B}^{jk}  P_0^k \,\approx\,0
\end{equation}

\noindent in the $\epsilon \rightarrow 0$ limit this gives the finite result

\begin{equation}
\Big( \partial_\sigma X^i \,-\,2\pi \alpha^\prime\,
 M^{-1}_{ij}  {\cal B}_{jk} P_0^k \Big)\vert_{_{\sigma =0}} \,\approx\,0
\end{equation}

\noindent where $ M_{ij} \,=\, \delta_{ij} \,-\,{\cal B}_{ik} {\cal B}_{kj} \,$. 
This equation corresponds to the first of the  constraints of eq. (\ref{e1}) 
expressed in terms of the symplectic variable  $P_0^k $.

For the second constraint we find  

\begin{equation}
\Omega_2^i \,=\,
- \epsilon \lbrace 2 \pi \alpha^\prime\,P_{oi}
\,+\, {\cal B}_{ij} {( X^j_1 - X^j_0 ) \over \epsilon }\,\rbrace\,\approx\,0\,.
\end{equation}

\noindent So, this actually gives no constraint  as $\epsilon \rightarrow 0$. 

Now for the third constraint we find

\begin{equation}
\Omega_3^i \,=\,
{( X^i_N - X^i_{N-1} ) \over \epsilon } \, - \, 2 \pi \alpha^\prime\,
 M^{-1}_{ij}{\cal B}^{jk}  P_{N-1}^k \,\approx\,0
\end{equation}

\noindent that leads, when $\epsilon \rightarrow 0$, to the finite result

\begin{equation}
\Big( \partial_\sigma X^i \,-\,2\pi \alpha^\prime\,
 M^{-1}_{ij}  {\cal B}_{jk} P_N^k \Big)\vert_{_{\sigma = \pi}} \,\approx\,0\,\,,
\end{equation}

\noindent that corresponds to the second boundary condition of eq. (\ref{e1}). 
The last constraint will give no contribution when $\epsilon \rightarrow 0$

\begin{equation}
\Omega_4^i \,=\,
- \epsilon 2 \pi \alpha^\prime\,P_N^i\,\approx\,0\,.
\end{equation}

In order to  incorporate the constraints $\Omega_1^i\,$ and $\Omega_3^i\,$ into 
the symplectic formalism we introduce Lagrange multiplier variables 
 $\lambda_1^i$ and $\lambda_3^i$ and add  a new term to the Lagrangian

\begin{equation}
L_1\,=\, L_0 \,+\,
\dot\lambda_1^i \lbrace
 {( X^i_1 - X^i_0 ) \over \epsilon } \, - \, 2 \pi \alpha^\prime\,
 M^{-1}_{ij}{\cal B}^{jk}  P_0^k \,\rbrace
\,+\,\dot\lambda_3^i \lbrace
{( X^i_N - X^i_{N-1} ) \over \epsilon } \, - \, 2 \pi \alpha^\prime\,
 M^{-1}_{ij}{\cal B}^{jk}  P_{N-1}^k \,
\rbrace
\end{equation}

Now the symplectic matrix becomes

\begin{equation}
\bordermatrix{&X_0^j&P_0^j&...&X_n^j&P_n^j&...&X_N^j&P_N^j&\lambda_1^j
&\lambda_3^j\cr
X_0^i&0&-\epsilon\delta^{ij}&...&0&0&...&0&0&\delta_{ij}/\epsilon&0\cr
P_0^i&\epsilon\delta^{ij}&0&...&0&0&...&0&0&\Gamma^{ij}&0\cr 
...& ...&...&...&...&...&...&...&...&...&...\cr
X_n^i&0&0&...&0&-\delta^{ij}&...&0&0&0&0\cr
P_n^i&0&0&...&\delta^{ij}&0&...&0&0&0&0\cr
...& ...&...&...&...&...&...&...&...&...&...\cr
X_N^i&0&0&...&0&0&...&0&-\epsilon\delta^{ij}&0&-\delta_{ij}/\epsilon\cr
P_N^i&0&0&...&0&0&...&\epsilon\delta^{ij}&0&0&\Gamma^{ij}\cr
\lambda_1^i&-\delta_{ij}/\epsilon&\Gamma^{ij}
&...&0&0&...&0&0&0&0\cr
\lambda_3^i&0&0&...&0&0&...&\delta_{ij}/\epsilon&\Gamma^{ij}&0&0\cr
}
\end{equation}

\noindent where $\Gamma^{ij}\,=\, -2\pi\alpha^\prime \Big( M^{-1} {\cal B}\Big)_{ij}$.

As this symplectic matrix is non singular,  the elements of the inverse will be the 
corresponding commutators. The relevant ones are

\begin{eqnarray}
\lbrack X_0^i , X_0^j \rbrack &=& - { \Gamma^{ij}\over 2} \nonumber\\
\lbrack  X_n^i , X_n^j \rbrack &=&  0 \nonumber\\
\lbrack  X_N^i , X_N^j \rbrack &=&  { \Gamma^{ij}\over 2} 
\end{eqnarray}

\noindent where $ X_n^i $ represents a point that has a finite distance to 
 the end points  of the string. This result corresponds to the expected commutator 
for the string end points:

\begin{eqnarray}
\lbrack X^i(\sigma=0) , X^j (\sigma = 0 ) \rbrack &=& 
\pi\alpha^\prime \Big( M^{-1}
 {\cal B}\Big)^{ij} \nonumber\\
\lbrack  X^i (\sigma =\pi) , X^j (\sigma =\pi) \rbrack &=&  
-\pi\alpha^\prime \Big( M^{-1} {\cal B}\Big)^{ij} 
\end{eqnarray}

So, the non commutativity of a string attached to  a brane in the presence
of an antisymmetric tensor field is reproduced in a straightforward way.
It is important to stress the crucial role of our choice of symplectic variables
(\ref{SV}) where the factors of $1/ \epsilon$ lead to the incorporation of 
the boundary conditions in the quantization.

\vskip 1cm
Acknowledgements: The authors are partially supported by  CNPq., 
FAPERJ and FUJB (Brazilian Research Agencies).

\end{document}